\documentclass[twocolumn]{aastex7} 
\usepackage{menukeys}
\usepackage{gensymb}

\usepackage{hyperref} 
\makeatletter
\hypersetup{
    colorlinks=true,  
    linkcolor=blue,  
    citecolor=blue,   
    filecolor=blue,    
    urlcolor=blue
}
\makeatother

\begin{document}

\title{Spatio-Temporal Evolution of a Steady Transition Region Downflow above a Sunspot}

\correspondingauthor
{S. Krishna Prasad}
\email{krishna.prasad@aries.res.in}

\author[0009-0003-7185-5727]{Bablu Mandal}
\affiliation{Aryabhatta Research Institute of Observational Sciences, Manora Peak, Nainital-263001, Uttarakhand, India}
\affiliation{Department of Physics,
Indian Institute of Technology Roorkee,
Roorkee-247667, Uttarakhand, India}
\email{mandalbablu1997@gmail.com}  

\author[0000-0002-0735-4501]{S. Krishna Prasad}
\affiliation{Aryabhatta Research Institute of Observational Sciences, Manora Peak, Nainital-263001, Uttarakhand, India}
\email{krishna@aries.res.in}

\author[0000-0003-0819-464X]{Avijeet Prasad}
\affiliation{Statkraft AS, Lysaker, Norway}
\email{prasad.avijeet@gmail.com}

\author[0000-0001-8232-6626]{Ambrish Pandey}
\affiliation{Department of Physics,
Indian Institute of Technology Roorkee,
Roorkee-247667, Uttarakhand, India}
\email{ambrish.pandey@ph.iitr.ac.in}

\begin{abstract}
Super-Sonic Downflows (SSDs) are usually observed above sunspots as redshifted peaks in lower Transition Region (TR) spectral lines and can last from a few minutes to several hours. Utilising high-cadence, $\sim$3 hour long repeated-raster mode observations from Interface Region Imaging Spectrograph (IRIS), we investigate the spatio-temporal dynamics of such a downflow event above NOAA active region 12135. Triple Gaussian fitting to TR spectral lines, along with Non-Force Free Field (NFFF) extrapolation and Differential Emission Measure (DEM) analysis, is performed to study this event. Downflow is initially observed in both the umbral and penumbral regions. Over time, the penumbral downflow fades, while the umbral downflow strengthens and appears as a single structure initially but eventually splits into two spatial segments. The calculated average properties, including downflow speed, electron density, and mass flux for this event, align with established studies. Simultaneous evolution in overlying coronal loops is also observed using imaging data from the Atmospheric Imaging Assembly (AIA). DEM analysis of the coronal emission over the downflow region indicates corresponding changes in cool ($\log_{10}[T/\mathrm{K}] \in [5.5, 5.6]$) coronal plasma. As suggested by recurrent transient brightenings observed in the nearby region in coronal channels and the corresponding fan-spine magnetic topology identified from NFFF extrapolations, magnetic reconnection occurring in this region is proposed to be the cause of this evolution. Our analysis revealed that changes in the morphology of TR downflows are closely associated with evolution of overlying coronal loops.
\end{abstract}


\section{Introduction} 
Sunspots, the manifestation of strong magnetic fields, appear as dark patches on the solar surface in white-light images \citep{1908_Hale_MagField,2003_Solanki_Sunspot}. Each spot consists of two parts, the umbra (the central dark region) and the penumbra (the surrounding less dark region). Above the sunspots, a plethora of dynamic phenomena, including oscillations and waves (see \citet{2015_Khomenko_Sunspot_Oscillations_Waves} for a detailed review), and plasma flow events like Evershed Flow \citep{1909_Evershed_EF} and Inverse Evershed Flow \citep{1913_John_IEF, 2022_Avijjet_Prasad_Inverse_Evershed_Flow} are observed throughout the layers of the Solar atmosphere. Additionally, Super-Sonic Downflows (SSDs) are also observed above sunspots as redshifted peaks in the lower TR spectral lines \citep{1982_Dere_SSD, 1982_Nicolas_SSD, Brynildsen2001, Brynildsen2004}. 

As noted in earlier Solar Ultraviolet Measurements of Emitted Radiation (SUMER: \citet{1995_Wilhelm_SUMER}) observations by \citet{Brynildsen2001}, these events can be identified as ``dual flows'', characterised by two distinct Gaussian peaks of the same spectral line. These authors reported that the component with the smallest wavelength shift displays 3-minute oscillations at subsonic speeds, while the other component maintains a supersonic speed without any significant oscillations within the largest umbra of AR 8989 and AR 8998. Based on Interface Region Imaging Spectrograph (IRIS: \citet{2014_DePontieu_IRIS}) raster observations (moving the slit with various numbers of steps and step sizes to scan a region of interest), a detailed statistical study by \citet{2018_Samanta_SSDs} reported that SSDs appeared in $48$ out of $60$ (80\%) sunspots, 28 of which occurring in the penumbrae, 4 in the umbrae and 16 in both regions. Sometimes, TR downflows also leave a signature in the Chromospheric spectral lines \citep{Chitta2016, 2018_Samanta_SSDs, 2020_Nelson_SSD}, suggesting there may be a potential link between TR downflows and observed chromospheric downflows within sunspots \citep{2022_Sowmya_chromosphericSSD}. 
\citet{2020_Nelson_SSD, 2020_Nelson_raster_downflows} demonstrated that SSDs are localised phenomena, occurring over a few arc-seconds, and displaying strong variability over minutes to hours. Their frequent disappearance at the same positions points to the possibility that sunspots may host more SSDs than previously estimated by \citet{2018_Samanta_SSDs}, as raster scans may not adequately capture all instances of downflows.\\
\begin{figure*}
    \centering
\includegraphics[width=\textwidth]{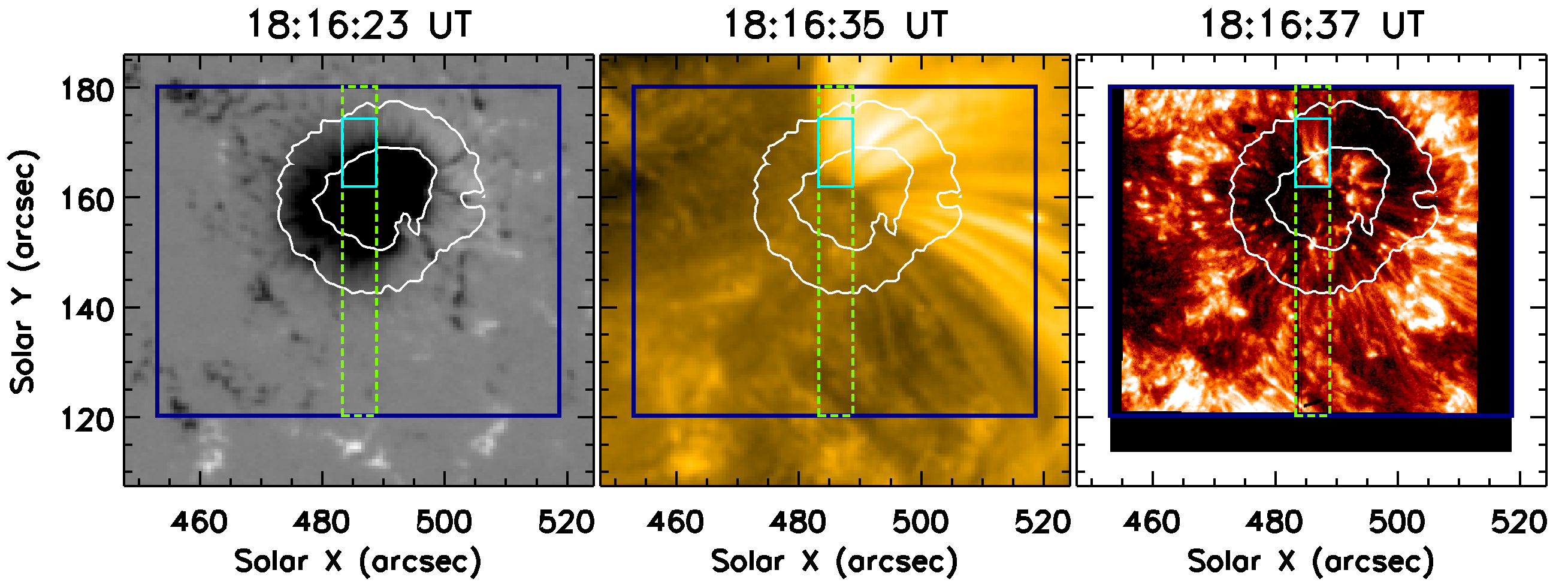}\\[-1.0ex]
    \caption{Contextual image of a sunspot within AR $12135$ as observed on August $13,2014$ extracted from SDO/HMI line of sight Magnetogram saturated at $\pm800~\mathrm{G}$ (left), corresponding closest image from SDO/AIA $171$\AA\ channel (middle) and IRIS/SJI $1400$\AA\ channel (right), exhibiting IRIS SJI FOV (thick solid navy box), IRIS Raster FOV (dotted green box) and SSD location (thin cyan box). Following a similar approach mentioned in \citet{Sanjay_2024} and \citet{KrishnaPrasad_2017}, the umbra-penumbra boundary (inner) as well as the penumbra boundary (outer) contours in white are defined by a percentage of median intensity in quiet sun regions, using time-average HMI continuum images throughout the observation and overlaid in each panel.}
    \label{figaia171_hmicontinuum}
\end{figure*}
Based on their spectral signatures, these events can be divided into two distinct categories, Bursty SSDs \citep{Kleint2014} and Steady SSDs \citep[][]{Straus2015, 2020_Nelson_SSD, 2020_Nelson_raster_downflows, Chen2022}. Bursty SSDs show extremely broad line profiles with apparent downflow speeds reaching up to $200$ $\mathrm{km/s}$ at the red wings of lower TR lines (Si IV $1394$ \AA, Si IV $1402$ \AA\ and C II $1336$ \AA) as well as Chromospheric lines (Mg II h $2803$ \AA\ and Mg II k $2796$ \AA). Due to their high intermittency, they have usually been associated with coronal rain \citep{2020_Antolin_CoronalRain, Ishikawa2020}. Steady SSDs generally show two well-separated line profiles in the red wings, with an apparent speed of around 100 $\mathrm{km/s}$ and can have lifespans of several hours. Due to their high mass-flux ($\sim10^{-7} \, \text{g} \, \text{cm}^{-2}\, \text{s}^{-1} $), often their origins are linked to siphon flow, caused by pressure differences between the loop footpoints \citep{1980_Cargil_Priest_SiphonFlow, 1981_Noci_SiphonFlow}. More recently, \citet{Chen2022} suggested that these 
long-lived steady SSDs can also have coronal origin, due to plasma overflow resulting from a reconnection-facilitated coronal condensation in newly formed magnetic dips.

A sit and stare (fixed slit mode) observation can capture the detailed temporal behaviour of these events \citep{Straus2015, Chitta2016, 2020_Nelson_SSD}, in contrast, a single raster scan only provides two-dimensional morphological information \citep{2018_Samanta_SSDs, 2020_Nelson_raster_downflows}. Although \citet{2020_Nelson_raster_downflows} attempted to bridge this gap using repeated raster mode observations from IRIS, a high-cadence, long-duration study is still lacking in the existing literature, which is crucial for a complete understanding of the spatio-temporal evolution of these flows.\\ 
In this article, we investigate the evolution of a long-lived SSD in AR 12135 by utilising nearly 3-hour-long repeated raster mode IRIS observations. The aim of this study is to better understand the spatio-temporal evolution of these sunspot-specific flows in the solar atmosphere. The observational data for this study are described in Section~\ref{sec:Observations}, and its analysis is provided in Section~\ref{sec:Data Analysis}. Details about NFFF extrapolation are introduced in Section~\ref{sec: Magnetic Field Extrapolations}. The obtained results are presented in Section~\ref{sec:Results}, followed by a discussion as well as a concise summary in Section~\ref{sec:Discussions}.
\section{Observations} \label{sec:Observations}
For this study, data are obtained from the space-based IRIS satellite, which can capture images through the Slit-Jaw Imager (SJI) and collect simultaneous spectra using three spectral channels: a near-ultraviolet (NUV) channel covering $2782$–$2835$ \AA\, and two far-ultraviolet (FUV) channels at $1332$–$1358$ \AA\ (FUVS) and $1389$–$1407$ \AA\ (FUVL). The NUV channel covers two strong spectral lines formed in the chromosphere (Mg II h $2803$ \AA\ and Mg II k $2796$ \AA), and the FUVL channel can record several TR lines, such as the Si IV $1394/1402$ \AA\ and O IV $1399/1401$ \AA\ lines. SJI has a spatial resolution of $0.33''$ in the NUV band and $0.4''$ in the FUV band. In the FUV channels, the effective spectral resolution is $26$ m\AA, compared to $53$ m\AA\ for the NUV channel \citep{2014_DePontieu_IRIS}.\\
The data we use are a Level 2 data\footnote[1]{The IRIS data and cutout co-aligned AIA data can be accessed here: \  \url{https://iris.lmsal.com/search/}}, processed with dark current subtraction, flat fielding, and geometrical corrections, focused on the sunspot of AR $12135$, collected in repeated raster mode on August 13, 2014, from 18:16:37 to 21:10:07 UT, consisting of 200 rasters. Field of view of each raster is $(5'' \times 60'')$, and SJI is $(65'' \times 60'')$. SJI images are taken in two filters ($2796 \, \text{\AA} \, \text{and} \, 1400 \, \text{\AA}$) at a cadence of $13$ s, while raster cadence is $52$ s. For rasters, the step-size is $0.350''$, and the pixel scale along the slit is $0.166''$, whereas the pixel size of slit jaw images, is $0.166'' \times 0.166''$. There are a total $8$ spectral windows in this data set, $1335.71$ Å (C II $1336$), $1349.43$ Å (Fe XII $1349$), 1355.60 Å (O I 1356), 1393.78 Å (Si IV 1394), 1402.77 Å (Si IV 1403), 2796.20 Å (Mg II k 2796), 2814.65 Å (2814) and 2832.93 Å (2832).

In order to gain knowledge about multi-temperature plasmas and the magnetic field configuration, as well as to provide contextual images, we supplemented the IRIS observations with data from Solar Dynamics Observatory \citep[SDO; ][]{2012_Pesnell_SDO}, primarily from the Atmospheric Imaging Assembly \citep[SDO/AIA; ][]{2012_Lemen_AIA} and the Helioseismic and Magnetic Imager \citep[SDO/HMI; ][]{2012_Scherrer_HMI}. The HMI data are available for use in Level 1 strata\footnote[2]{The SDO/HMI data can be accessed here: \\  \url{http://jsoc.stanford.edu/ajax/exportdata.html}}, where they have already been processed to remove bad pixels and spikes. However, in this work, we use Level 1.5 HMI data obtained using \texttt{aia\_prep.pro} routine available in the SolarSoft library \citep{1998_Freeland_SolarDataAnalysis} of \texttt{Interactive Data Language} (IDL). This routine aligns the plate scale of the HMI data with that of AIA, such that both datasets have a pixel scale of $0.6''$. It also corrects the rotation of the Sun with respect to the image plane, thereby aligning Solar North with image North. Contextual image of AR $12135$ combining these datasets is shown in Fig.~\ref{figaia171_hmicontinuum}. The HMI line-of-sight Magnetogram revealed that our region of interest within AR $12135$, is a sunspot of negative polarity. AIA $171$ image highlighted the emanating coronal fan loops from the umbral periphery, while the SJI $1400$ image shows the loop footpoints in TR.
\section{Data Analysis} 
\label{sec:Data Analysis}
Since, the FOV of each raster is small ($5''$) in X-direction, we verify the coalignment of individual rasters with respect to corresponding SJI images, but did not find any significant shift. We also align IRIS and SDO data by visually matching features in the SDO/AIA $1600$ \AA\ and IRIS/SJI $1400$ \AA\ channels.

For absolute wavelength calibration, in TR spectra, neutral or singly ionised lines from the photosphere and the chromosphere can be used \citep{1991_Hassler}. IRIS FUV spectral windows contain such lines, for e.g., Fe II $1392.8$, S I $1401.514$ and O I $1355.598$. \citet{2019_Ghosh_CLV_TR} points out that the Fe II $1392.8$ line is unsuitable for absolute wavelength calibration due to the asymmetry in its blue wing. On the other hand, the neutral S I $1401.514$ \AA\ line in the Si IV $1402$ spectral window is blended with a nearby line. Therefore, we choose neutral O I $1355.598$ \AA\ line for final wavelength calibration for the FUVL channel. 


\begin{figure*}[t]
    \centering
\includegraphics[width=0.492\linewidth]{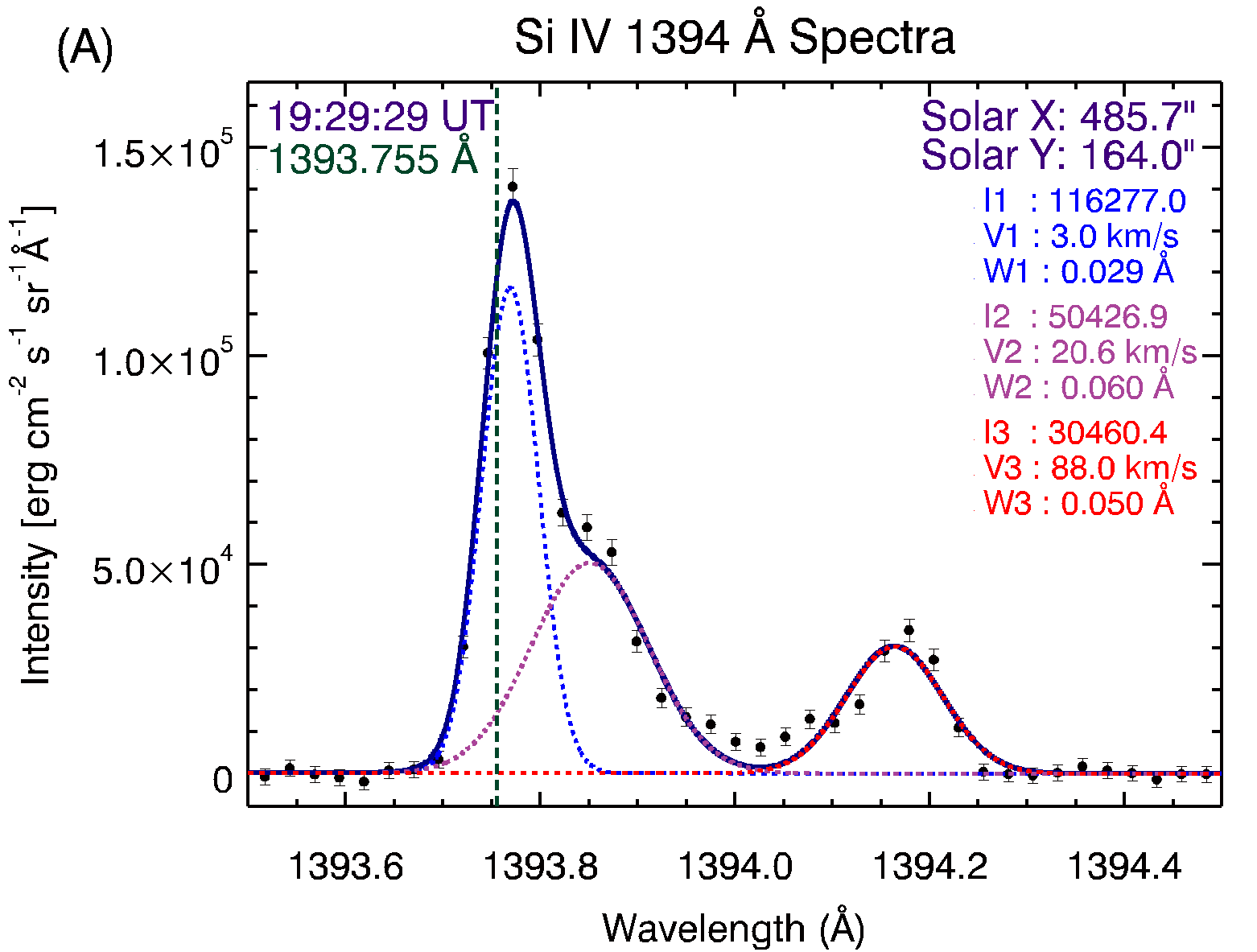}\hspace{0.01\linewidth}
\includegraphics[width=0.492\linewidth]{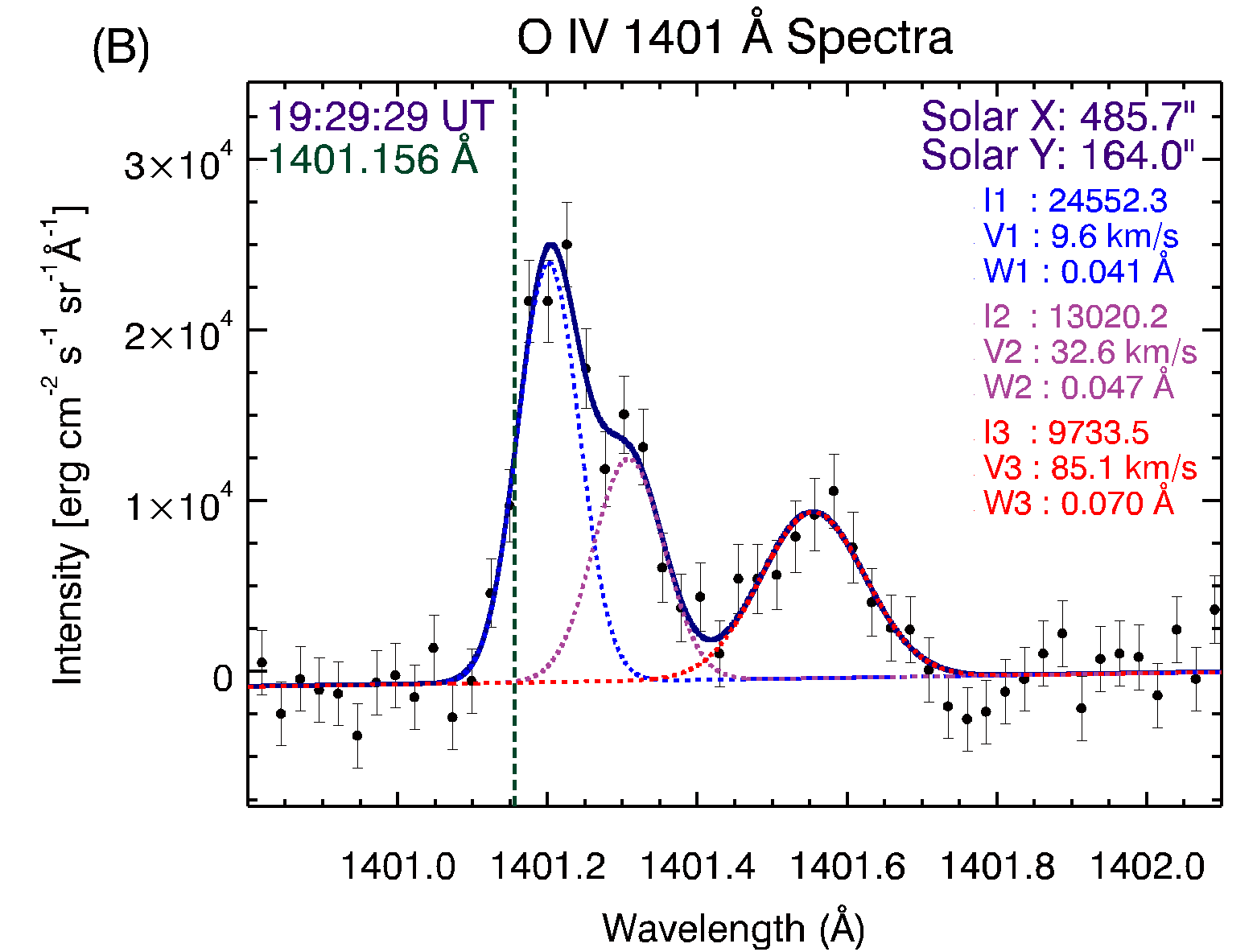}
\caption{Sample spectra for Si IV 1394 and O IV 1401 at a pixel location where downflows are observed during raster time 19:29:29 UT. In both panels, a dotted vertical line (green) marks the laboratory wavelength of the respective lines, with values referenced from \citet{2017_Tian_TR}. Individual Gaussian 1 (blue), Gaussian 2 (red) and Gaussian 3 (lavender) components are displayed in dotted curves from shorter to longer wavelength, respectively, while the combined Gaussian fit is shown in a solid curve (navy).}
\label{figspectralfitting}
\end{figure*}

 To derive the spectroscopic parameters of downflows, we have selected the Si IV $1394$ and $1403$ windows, as these windows contain Si IV $1393.755$ and $1402.77$ \AA\, the two strongest TR lines in the IRIS Spectral range and also density sensitive O IV $1399/1401$ Å doublet lines that are useful to determine the electron density of the TR plasma \citep{1992_Dwivedi_OIVDensity}. Prior to spectral fitting, radiometric calibration is performed using \texttt{iris\_getwindata.pro} routine with /calib and /perang keywords. To improve the signal-to-noise ratio, spectra are binned over two pixels in the Y direction, using \texttt{eis\_bin\_windata.pro}, resulting in a final pixel scale along Y of $0.333''$. After carefully checking changes in spectral shape over time, instead of traditional double Gaussian fitting, we performed triple Gaussian fitting to Si IV and O IV spectral lines on each pixel (using the procedure \texttt{iris\_auto\_fit.pro} available in the SolarSoft library) as depicted in \autoref{figspectralfitting}. A few rasters (9 out of 200) are contaminated due to a shower of charge particles hitting the detectors. Consequently, spectral fitting in those rasters is poor. So, we excluded those raster frames (121-129) from further analysis when calculating average properties (downflow intensity, downflow speed, electron density and mass flux) for this event. Certain constraints are imposed on the fitting parameters to obtain the best fits. The line intensities are set to be greater than zero, and the line widths are restricted to lie between $29$ mÅ and $150$ mÅ, a range similar to that applied in \cite{2020_Tripathi_Opacity}. For the Si IV lines at $1394$ \AA\ and $1402$ \AA, as well as the O IV lines at $1399$ \AA\ and $1401$ \AA, the peak of the first Gaussian is constrained to lie within $\pm20~\mathrm{km/s}$ of the respective rest wavelength. The second Gaussian component, constrained to a range of $20$ to $40~\mathrm{km/s}$, is introduced to account for a smaller peak that sometimes appears adjacent to the first Gaussian component. The third Gaussian's Peak is set in between $40$ to $150~\mathrm{km/s}$, based on the similar values reported in \cite{2018_Samanta_SSDs}. 
 
The downflow regions are visually identified by constructing images from peak intensities in the downflow component alone and isolated using a certain intensity threshold ranging from 3000 to 20000 in \(\mathrm{erg\ cm^{-2}\ s^{-1}\ sr^{-1}}\) \AA\(^{-1}\). Using a trial and error approach, this selection minimises noise, allowing us to derive the average properties of these regions for each spectral line and study their time-dependent changes. The average downflow intensity and downflow speed within these selected contours are estimated for all rasters (excluding 121-129). The corresponding errors are derived, by applying the propagation of error formula from the original 1$\sigma$ fitting errors.

The electron density in TR is calculated from the intensity ratio of the O IV $1399$ and $1401$ line pair \citep{1992_Dwivedi_OIVDensity, 2015_Young_IRIS, 2016_Polito_density_OIV, 2024_Liang_densityOIV}. The theoretical relations between intensity ratios of line pairs and corresponding electron densities are taken from the CHIANTI database \citep{1997_Dere_CHIANTI}. During this density estimate, the potential blend between the O IV $1401.16$ \AA\ line and the S I $1401.51$ \AA\ line is ignored as this is typically weak within sunspots \citep{2002_Keenan_OIV_SIV}. From the ratio of fitted intensities of these lines within the previously-mentioned contour values, average electron densities are estimated for each raster. Errors in density are calculated from estimated photon noise, which is then converted to density values. 

In order to assess multi-thermal coronal plasma in our region of interest, we have performed Differential Emission Measure (DEM) analysis, which quantifies plasma along the line of sight (LOS) as a function of temperature. The emission measure (EM), defined as $\text{EM} = \int_{T} \ DEM(T) \, \mathrm{d}T$ \ [ units: $\mathrm{cm}^{-5}] $, essentially represents the amount of plasma integrated along the LOS per unit area for a specific temperature range. To compute DEM, we have employed the new sparse inversion method \citep{2022_Li_Su_DEM2,2018_Su_DEM}\footnote[3]{Latest Sparse inversion code (version 1.001) can be downloaded from here: \url{http://paperdata.china-vo.org/yang.su/DEM/sparse_em_v1.001_ys.zip}}, a modified version of the approach described by \citet{2015_Cheung_DEM}. This method can accurately retrieve thermal plasma distribution within a temperature range of approximately $0.3$ to $30$ MK. The cutout AIA data provided alongside IRIS data in six SDO/AIA channels 94 \AA{} (characteristic temperature, $\log[T/\mathrm{K}]\sim6$), 131 \AA{} ($\log[T/\mathrm{K}] \sim 5.6$ and $7.0$), 171 \AA{} ($\log[T/\mathrm{K}] \sim 5.8$), 193 \AA{} ($\log[T/\mathrm{K}] \sim 6.2$ and $7.3$), 211 \AA{} ($\log[T/\mathrm{K}] \sim 6.3$), 335 \AA{} ($\log[T/\mathrm{K}] \sim 5.9$ and $6.4$) are utilized for this analysis. We have performed our inversion on a temperature grid spanning $\log_{10}(T/\text{K}) \in [5.5, 7.6]$ with a step size of $\Delta \log_{10}(T/\text{K})=0.05$. For a particular timestamp, data for each AIA channel are normalised by the corresponding exposure time. The resulting exposure-normalised 6 images are arranged into a 3D data cube, represented as $map\_in$ = $[x, y, \lambda]$, where $\lambda$ corresponds to the wavelength channels: 94, 131, 171, 193, 211, 335 in this sequence. This data cube is provided in  \texttt{get\_em\_from\_aia.pro} routine as input to estimate DEM. Finally, we have estimated the EM-weighted mean temperature for the  whole temperature range as follows
$\langle T \rangle_m = \frac{\sum \mathrm{EM} \times T }{\sum \mathrm{EM}}$. This process is repeated for each timestamp to capture the overall thermal evolution in the Corona.   
\section{Magnetic Field Extrapolations}
\label{sec: Magnetic Field Extrapolations}
 In order to understand the magnetic connectivity in the vicinity of AR 12135, we have performed magnetic field extrapolations using SDO/HMI vector magnetogram data. Traditionally, coronal magnetic field extrapolation techniques assume force-free conditions (Lorentz force = 0) to derive the magnetic field, using the photospheric magnetic field as the boundary condition \citep {2021_Wiegelmann_ForceFreeExtrapolations, Regnier2013}. However, in the lower solar atmosphere (the photosphere and the chromosphere) \citep{Metcalf_1995, 2001_Gary_PlasmaBeta}, as well as in scenarios involving high-speed plasma flow, force-free conditions may not hold. So, we employed the Non-Force-Free Field (NFFF) model\footnote[4]{NFFF code can be found here: \url{https://github.com/AvijeetPrasad/MDR-NFFF}}, a well-established method in the literature \citep{2018_APrasad,2020_APrasad,2023_APrasad}, to understand the underlying magnetic field configuration of AR 12135. Based on techniques developed by \citet{2008_HU_Dasgupta_NFFF} and \citet{2008_HU_MDR, 2010_HU_NFFF}, this model relies on the Minimum Dissipation Rate (MDR) principle. According to this principle, a two-fluid plasma, consisting of ions and electrons, naturally evolves towards a state that minimizes the total dissipation rate from both resistive and viscous effects, while conserving generalised helicities \citep{2007_Bhattacharyya_MDR}. Consequently, the magnetic field follows a double curl Beltrami equation \citep{1998_Mahajan_doublecurlBeltramieqn} and can be expressed as a sum of a potential field (zero twist along the magnetic field lines) and two linear force-free components (constant twist along the magnetic field lines), each satisfying $\nabla \times \mathbf{B}_i = \alpha_i \mathbf{B}_i$. By gradually decreasing the average difference between the observed ($\mathbf{B}_{t}$) and computed transverse fields ($\mathbf{b}_{t}$) on the photospheric boundary through iterations, these constant twist parameters are derived, following this metric \citep{2010_HU_NFFF, 2020_APrasad}{\large
\begin{equation}
E_n =
\frac{
\sum_{i=1}^{M} 
\left| \mathbf{B}_{t,i} - \mathbf{b}_{t,i} \right|
\, \times \left| \mathbf{B}_{t,i} \right|
}{
\sum_{i=1}^{M} 
\left| \mathbf{B}_{t,i} \right|^{2}
},
\end{equation}
}
where $M$ represents the total number of grid-points located on the photospheric boundary plane.

From SDO/HMI full-disk vector magnetic field data (\texttt{hmi.B\_720s} data series), initially a small cutout region of $(600 \times 600)$ pixels, centred at Carrington longitude 131\degree\ and Carrington latitude 16\degree, is extracted and then remapped onto the Lambert cylindrical equal-area (CEA) projection. The dataset has a pixel scale of $0.5''$ and a temporal cadence of 720 seconds. The final NFFF extrapolation is performed on this projected data within a computational domain defined by a grid of $(600 \times 600 \times 300)$ pixels along the $x$, $y$, and $z$ directions, respectively. This grid corresponds to a physical domain of $216\,\text{Mm}$ in the $x$ and $y$ directions, and $108\,\text{Mm}$ in the $z$ direction. We have utilised a vector magnetogram from SDO/HMI at 19:36:00 UT and performed NFFF extrapolation. Results from the extrapolation are described in \autoref{sec: Fan-Spine Topology near Sunspot and Magnetic-Reconnection behind splitting}.

\section{Results} \label{sec:Results}

\subsection{Evolution of downflows' morphology and its connection to coronal loop structures}
\label{sec: Spatio-Temporal Evolution of SSD}

\begin{figure*}
\centering
\includegraphics[width=\textwidth]{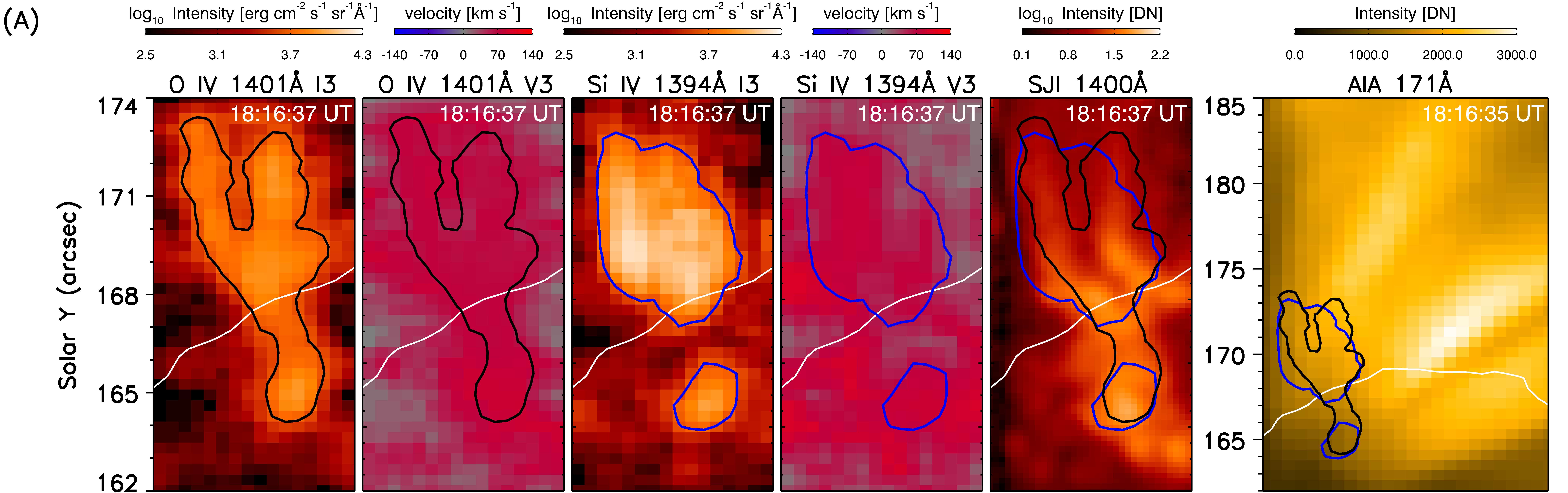}\\[-0.3ex]
  \includegraphics[width=\textwidth]{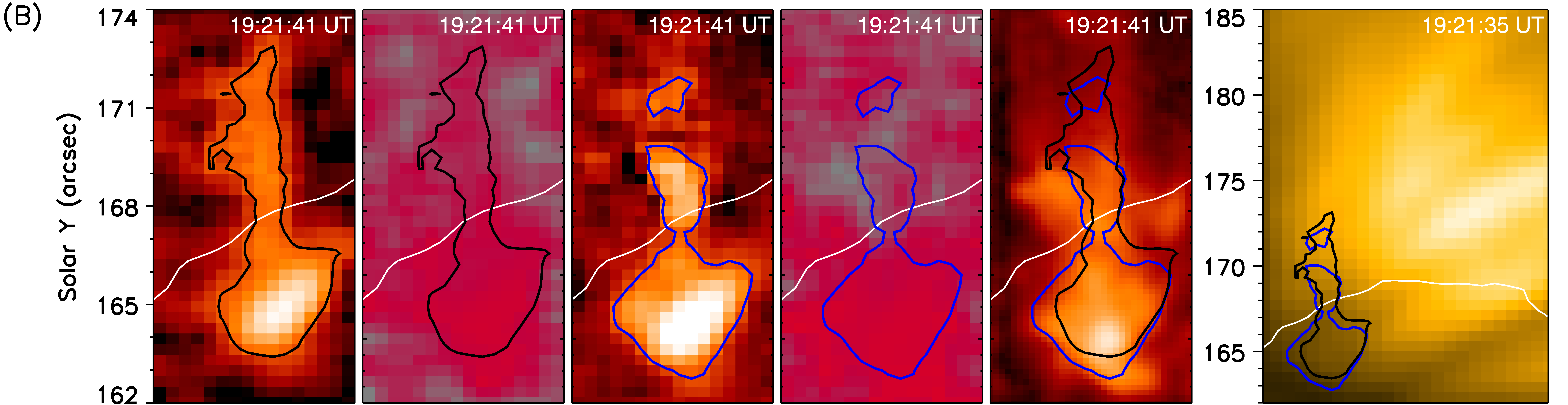}\\[-0.3ex]
  \includegraphics[width=\textwidth]{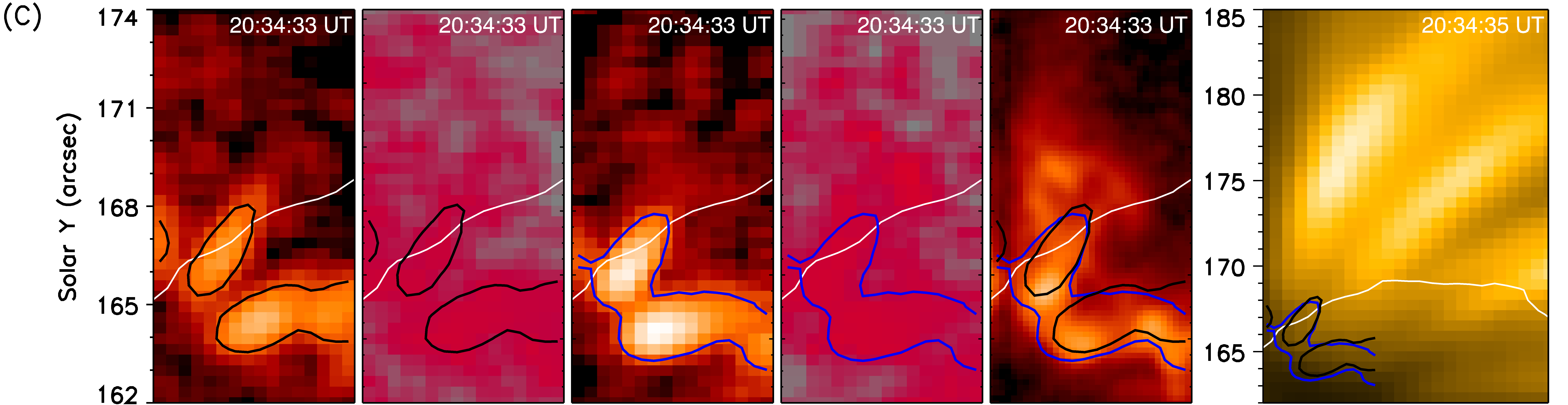}\\[-0.3ex]
  \includegraphics[width=\textwidth]{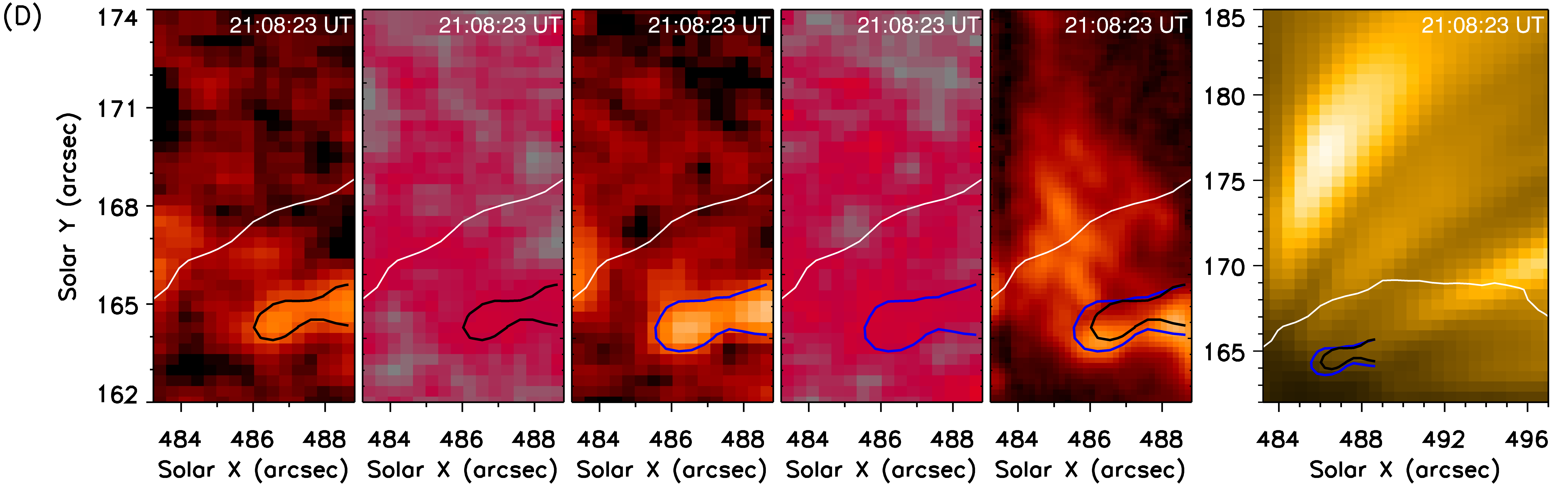}
  
\caption{Selected snapshots displaying spatio-temporal evolutions of the downflow event in O IV $1401$ \AA\ (downflow intensity), O IV $1401$ \AA\ (downflow speed), Si IV $1394$ \AA\ (downflow intensity), Si IV $1394$ \AA\ (downflow speed) and IRIS/SJI $1400$ \AA\ channel, along with the SDO/AIA $171$ \AA\ channel. Black and blue contours (with contour level equals to $\log_{10}(3.6)\ \mathrm{erg\ cm^{-2}\ s^{-1}\ sr^{-1}}$ \AA$^{-1}$) outline the downflow region observed in O IV $1401$ \AA\ and Si IV $1394$ \AA\ lines, respectively, while the white contour denotes the photospheric umbra-penumbra boundary. Refer to the text \autoref{sec: Spatio-Temporal Evolution of SSD} for more details.}
\label{fig:temporal_evolution_ssd}
\end{figure*}

After spectral fitting, we have obtained the fitting parameters (peak intensity, Doppler shift, linewidth) for each spectral line at each pixel in the raster FOV. We observed downflows in a small localised region within the sunspot. This region is marked by a cyan box in \autoref{figaia171_hmicontinuum}. The spatio-temporal evolution of downflow within that region is shown in different spectral lines from IRIS (Si IV $1394$ and O IV $1401$) as well as the imaging filters (IRIS/SJI $1400$ and SDO/AIA $171$) as seen in \autoref{fig:temporal_evolution_ssd}. Although we performed the analysis for both the Si IV $1394$ and $1403$ spectral lines, here we present only the intensity and velocity maps of the Si IV $1394$ line, as it is the strongest and exhibits evolution almost identical to the Si IV $1403$ line. The first and third columns in \autoref{fig:temporal_evolution_ssd}, contain emission only from the downflow component, while the second and fourth columns show the corresponding line of sight downflow speed. The fifth column contains both the main and downflow components and probably more, because it is from an imaging filter of wider bandpass, and the last column shows the evolution in the corresponding coronal structures.
\begin{figure*}
\centering
\includegraphics[width=\textwidth]{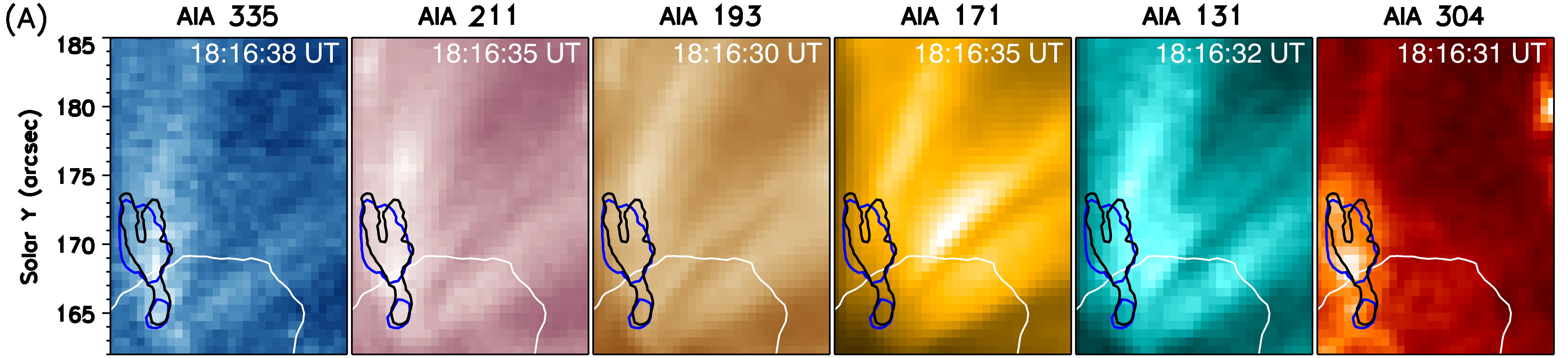}\\[-0.7ex]
  \includegraphics[width=\textwidth]{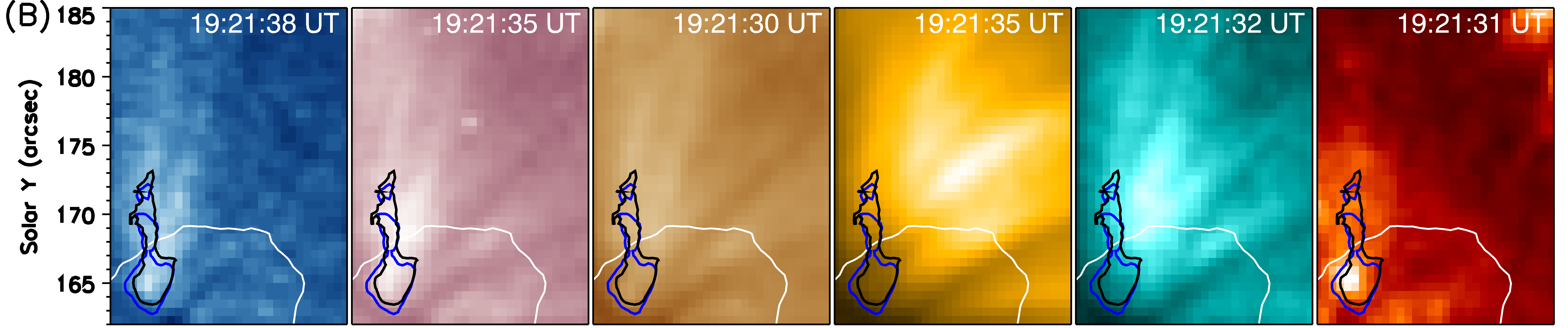}\\[-0.7ex]
  \includegraphics[width=\textwidth]{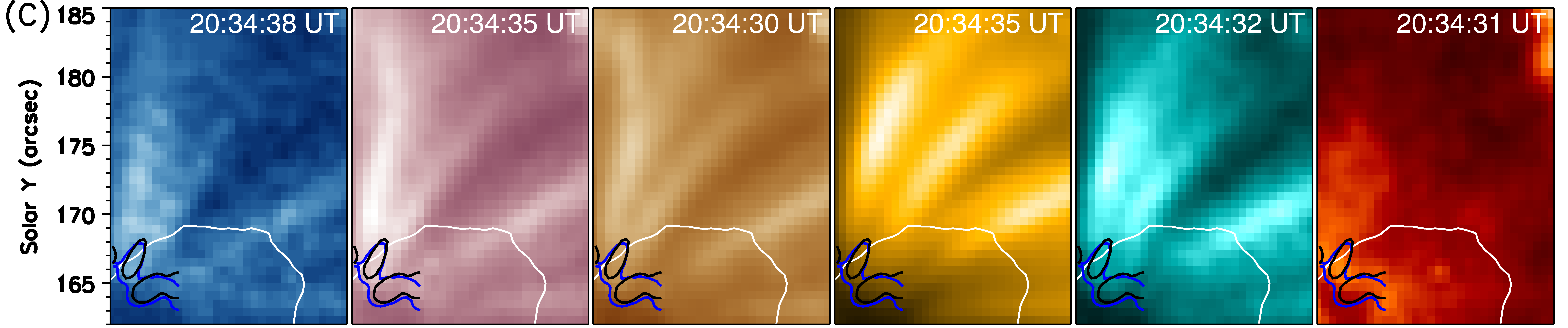}\\[-0.7ex]
  \includegraphics[width=\textwidth]{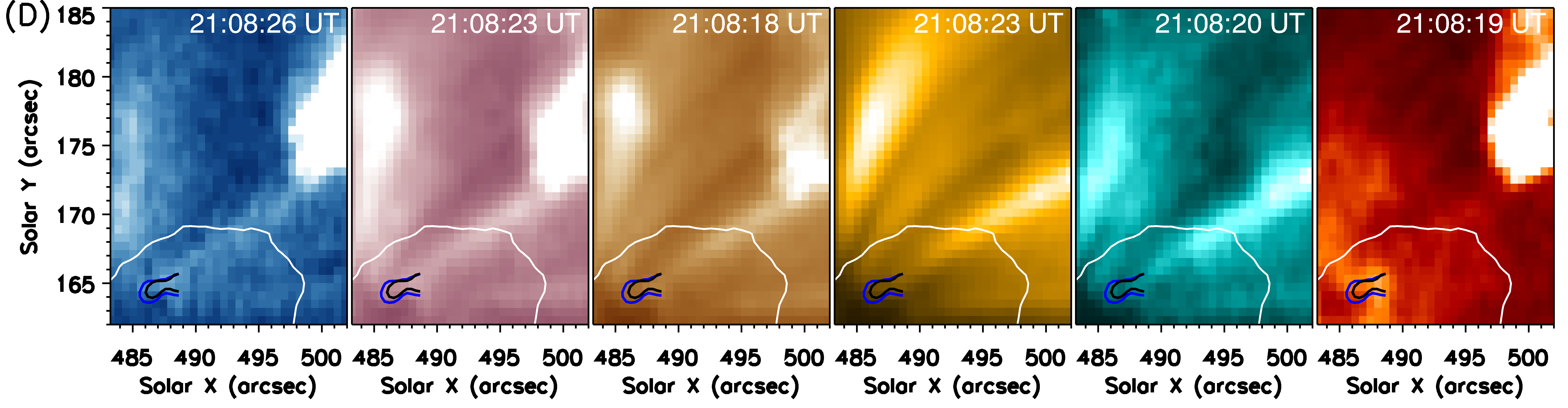}
    \caption{Structure of coronal loops across various AIA channels in the vicinity of the downflows. Different rows correspond to the data at 4 different times identified in \autoref{fig:temporal_evolution_ssd}. The blue/black contours highlight the downflow pattern observed in Si IV 1394 \AA/O IV 1401 \AA\ lines.}
\label{fig:temporal_evolution_ssd_aia}
\end{figure*}
\begin{figure*}
    \centering
\includegraphics[width=\linewidth]
{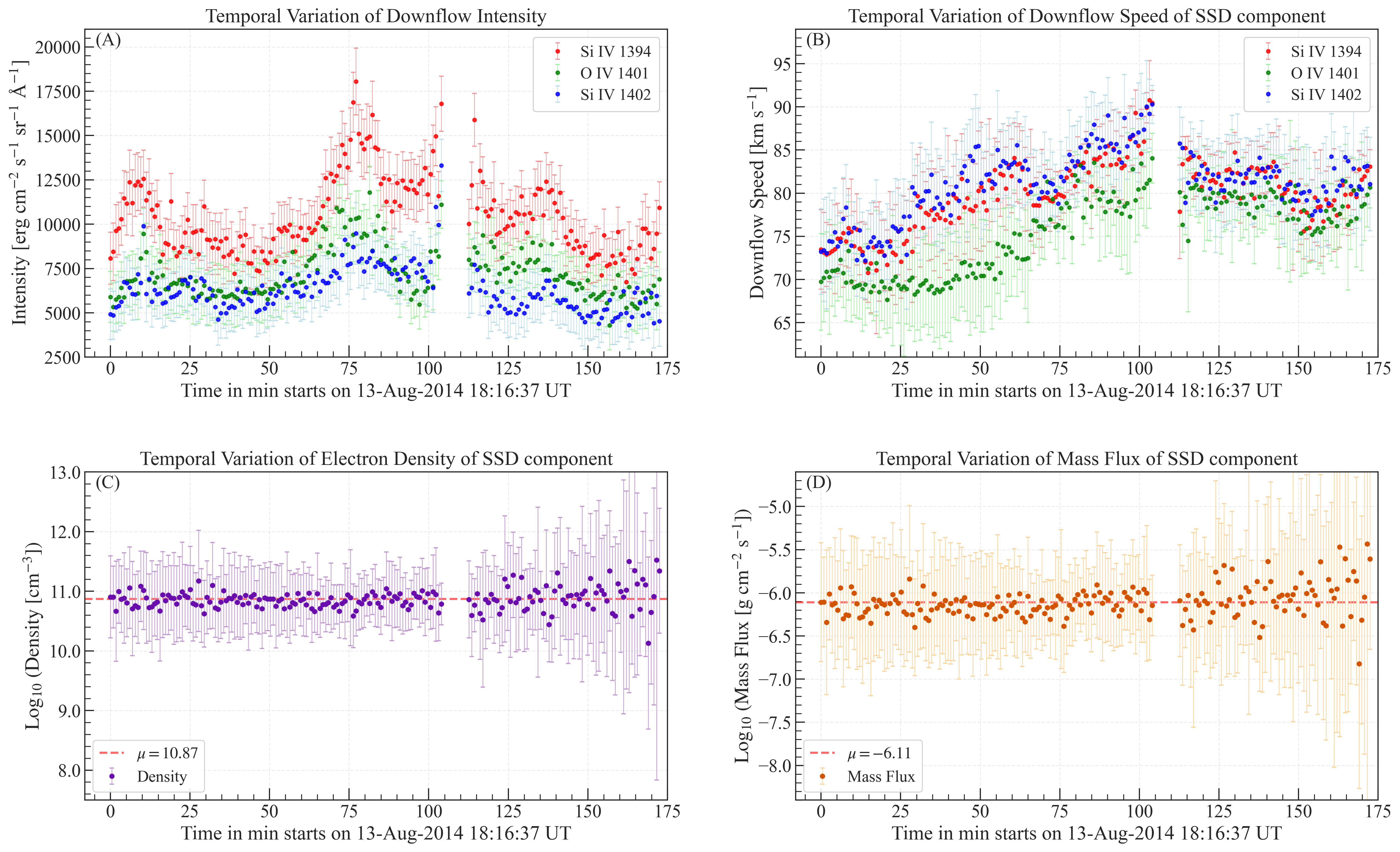} 
    \caption{Temporal variation of the characteristics of downflow, including average downflow intensity, their speed, electron density and mass flux, is presented here. The global mean ($\mu$) values of electron density and mass flux are indicated by a dashed red line in the respective panels.}
\label{figdownflowproperties}
\end{figure*}

\begin{figure*}
    \centering
\includegraphics[width=\textwidth]{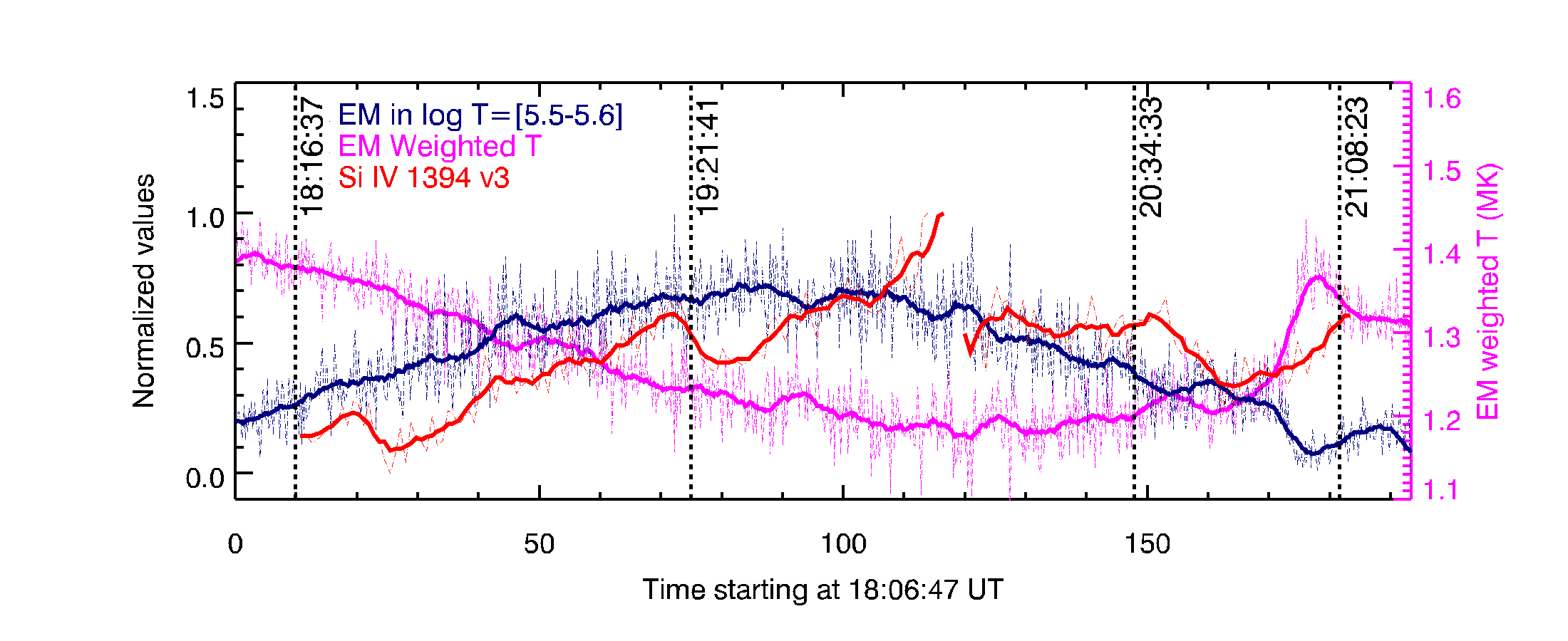}\\[-1.0ex]
    \caption{The average temporal variation of normalised EM in $\log_{10}(T/\mathrm{K}) \in [5.5, 5.6]$ range, along with normalized downflow speed for Si IV 1394 \AA\ channel (within the cyan box in \autoref{figaia171_hmicontinuum}), are shown. The variation in EM-weighted Temperature is also presented here. The four vertical dashed lines indicate timestamps corresponding to significant changes in the morphology of the downflow.}
    \label{fig:EM_LightCurves}
\end{figure*}
\begin{table*}
\caption{Comparison of SSD properties with previous research works.}
\label{ssds_properties_table}
\centering
\renewcommand{\arraystretch}{1.2}
\begin{tabular}{l l c c}
\hline\hline
Reference & Speed & $\log(\text{Density})$ & Mass Flux \\
 & (km s$^{-1}$) & (cm$^{-3}$) & (g cm$^{-2}$ s$^{-1}$) \\
\hline
\citet{Straus2015} & O IV 1401 = $82 \pm 8$& $10.6 \pm 0.25$ & $\sim 5 \times 10^{-7}$ \\
 & Si IV 1402 = $91 \pm 8$ & & \\
\hline
\citet{Chitta2016} & O IV 1401 $\sim$ 100 & -- & -- \\
 & Si IV 1402 $\sim$ 100 & & \\
 & Mg II 2796 $\sim$ 100 & & \\
\hline
\citet{2018_Samanta_SSDs} & O IV 1401 = $103 \pm 15$ & $10.17 \pm 0.69$ & $\sim 2 \times 10^{-7}$ \\
 & Si IV 1402 = $101 \pm 16$  & & \\
\hline
\citet{2020_Nelson_SSD} & O IV 1401 = 60 -- 108 & 9.6 -- 10.2 & $3.3 \times 10^{-8}$ -- $1.5 \times 10^{-7}$\\
 & Si IV 1394 = 57 -- 112\\
 & C II 1334 = 100 & & \\
\hline
\citet{Chen2022} & O IV 1401 $\sim$ 95 & $\sim 10.4$ & $\sim 5 \times 10^{-7}$ \\
 & Si IV 1402 $\sim$ 95 & & \\
 & Mg II 2796 $\sim$ 135 & & \\
\hline
This Work & O IV 1401 $\sim$ 85 & $\sim 10.87$ & $\sim 7.8 \times 10^{-7}$ \\
 & Si IV 1402 $\sim$ 90 & & \\ 
 & Si IV 1394 $\sim$ 90 & & \\
\hline
\end{tabular}
\end{table*}
\begin{figure*}
\centering
\includegraphics[width=\linewidth]{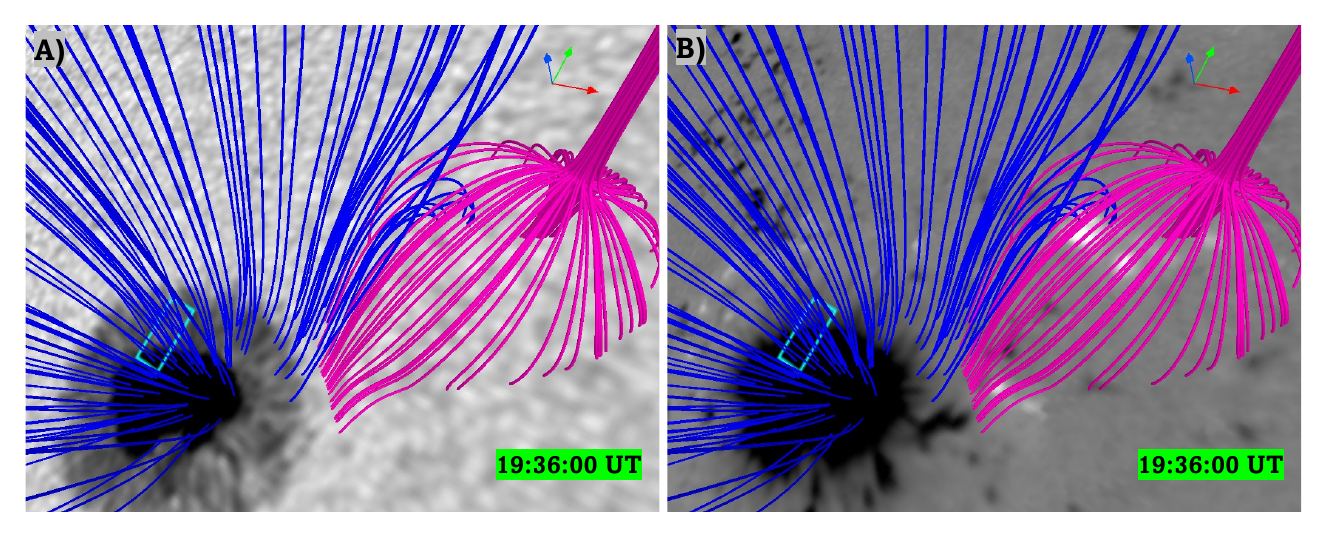}
\caption{NFFF extrapolation of AR 12135. Here, the magenta field lines highlight the fan-spine topology, while blue field lines indicate the ambient magnetic field. The red, green, and blue arrows point to the X, Y and Z directions, respectively. Panels (A) and (B) illustrate the NFFF extrapolated magnetic field lines at 19:36:00 UT, overlying an HMI Continuum image and an HMI LOS Magnetogram saturated at $\pm800~\mathrm{G}$, respectively. The cyan box represents the approximate location of downflow in CEA coordinates. Note that the images here are displayed with a viewing angle different from the plane of the sky. Above visualisation of extrapolations is achieved using VAPOR code \citep{VAPOR_Li_2019}.}
\label{fig:magextrapolations}
\end{figure*}
\begin{figure*}
\begin{interactive}{animation}{Updated_Movie/aia_rc_microflare_Updated.mp4}
\centering
\includegraphics[width=\linewidth]{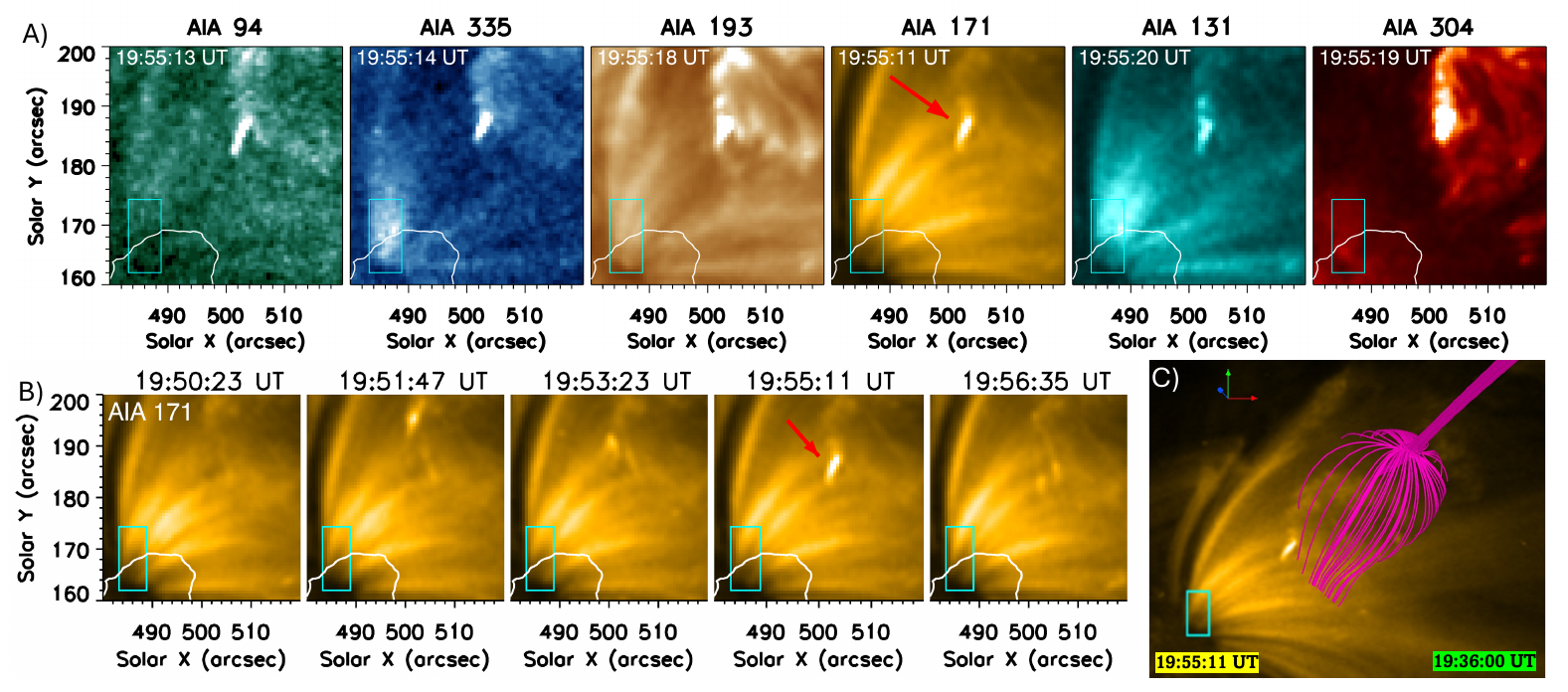}
\end{interactive}
\caption{Multi-wavelength appearance of a brightening near the sunspot as seen from different EUV channels of AIA is shown in panel (A). The white contour depicts the umbra-penumbra boundary and the cyan box represents the location of the downflow. The red arrow indicates the bright patch, and the evolution before and after the appearance of this patch is outlined in panel (B). Panel (C) displays the colocation of this brightening we see around 19:55:11 UT in the AIA 171 channel near the fan-spine topology identified from extrapolations at 19:36:00 UT. The cyan box here marks the approximate location of downflow in CEA projected coordinates. \\\textit{Panel (A) of this figure is available as a 32 s animation illustrating the temporal evolution highlighting recurring brightenings in this region.}}
\label{fig:rcAIA171}
\end{figure*}
We identify 4 different timestamps (18:16:37 UT, 19:21:41 UT, 20:34:33 UT, and 21:08:23 UT, shown in 4 rows in \autoref{fig:temporal_evolution_ssd}), when there is a notable change in the downflow behaviour. At the beginning of the observation at 18:16:37 UT, downflow is observed in both the umbral and penumbral regions, as shown in row (A). As time progresses, the intensity of the downflow in the penumbral region fades away, while the umbral downflows become more intense and appear as a single structure, as displayed in row (B). Meanwhile, in the SJI 1400 image, a part of the previously observed finger-like structure in the penumbral region has also disappeared, and the associated loop structure appears to fade away in the AIA 171 channel. The downflow region eventually appears to fragment into two parts, as depicted in row (C). At this stage, the structure observed in SJI 1400 appears to co-evolve, forming an inclined V shape($<$). Notably, the loop structures in 171\AA\ now become more apparent. By the time of observation at 21:08:23 UT, a single downflow component remains, as observed in row (D).

To further analyse the association between the change in the morphology of the downflow and the corresponding coronal loop evolution, we overlaid the downflow intensity contours of O IV $1401$\AA\ (black) and Si IV $1394$\AA\ (blue) lines across different channels of AIA in \autoref{fig:temporal_evolution_ssd_aia}. We excluded the 94 \AA\ channel, as the signal in this channel is poor. Because of this, the loop structure is not visible ( see Panel(A) of \autoref{fig:rcAIA171}). These AIA channels cover a wide temperature range from comparatively hotter $335$ \AA\ ($\log[T/\mathrm{K}] \sim 5.9$ and $6.4$) to cooler $304$ \AA\ ($\log[T/\mathrm{K}] \sim 4.7$) filter. At the beginning of the observation in row (A), both umbral and penumbral downflows are present, with loops visible in the respective locations across different channels. About one hour later, as the penumbral downflow diminishes, the loop that was previously visible in AIA 171 also fades from view at the same location. Interestingly, in row (C), the loops become more distinctive. A larger segment of downflow aligns with cooler structures observed in the AIA $304$ channel. This analysis reveals a close association between the evolution of coronal loops and the morphology of downflows.

\subsection{Estimation of average properties of the downflows}
\label{sec: Estimation of average properties for this downflow}
We derive the average properties of the downflows, including downflow speed, electron density and mass flux for this event. Using the methods outlined in \autoref{sec:Data Analysis}, the downflow speeds and densities are estimated, and the temporal variation of these parameters is presented in Panels (A) and (B) of \autoref{figdownflowproperties}. As may be noted, there is an acceleration in the downflow speed (speeds increasing from $\sim 73$ ($70$) km/s to $\sim 90$ ($85$) km/s over $103$ minutes for Si IV 1394 (O IV 1401)), followed by a slight deceleration observed in each spectral line. The formation temperatures for Si IV and O IV lines are approximately $T = 10^{4.88}$ K  and $T = 10^{5.17}$ K, respectively \citep{2018_Young_electrondensity}. Using these temperature values along with the formula for local sound speed, $c_s = 0.152 \sqrt{T}$ $\mathrm{km/s}$ \citep{2014_Priest}, we estimate the local sonic speeds to be around $42$ $\mathrm{km/s}$ for Si IV and $58$ $\mathrm{km/s}$ for O IV. So, these flows are apparently supersonic at the TR Temperature. On the other hand, the density remained nearly constant (within the uncertainties) throughout the observation, as shown in panel (C). The mass fluxes are usually computed using the formula, $ F_m = N_e m_p (N_H/N_e) V_D$, where, $N_e$ = number density of electron, $N_H$ = number density of hydrogen, $m_p$ = mass of proton, $V_D$ = speed of downflow. Previous studies \citep{Straus2015, 2018_Samanta_SSDs, 2020_Nelson_SSD} used a value of 0.83 for $N_H/N_e$, which we have adopted as well. The mass flux variation over time is estimated by integrating density values derived from O IV 1399/1401 line pairs and the downflow speed associated with the O IV 1401 line, as depicted in Panel (D). Additionally, \textit{\autoref{ssds_properties_table} }contains a comparison of these findings with results from previous studies.

\subsection{Results from DEM analysis}
\label{sec: Results from DEM analysis}

Within the observed downflow region (cyan box in \autoref{figaia171_hmicontinuum}), we measure average coronal emission measure with different bin sizes over the full range of $\log_{10}(T/\mathrm{K}) \in [5.5, 7.6]$ and examined if there is any correlation to TR downflow speed. As shown in \autoref{fig:EM_LightCurves}, Normalised EM in $\log_{10}(T/\mathrm{K}) \in [5.5, 5.6]$ range shows a similar evolution to the normalised Downflow speed (only Si IV 1394 v3 is shown), while the EM weighted mean Temperature light curve is showing an inverse relation.  This indicates a possible connection between TR downflow plasma and cool coronal plasma. It is worth noting that there are no correlated changes in emission measure at other temperature bins. While the increase in emission measure of the cool coronal plasma can be directly associated with increase in TR downflow intensity, the reason for corresponding increase in TR downflow velocity is not immediately obvious and requires advanced modelling studies.

\subsection{Magnetic-reconnection at fan-spine topology near sunspot}
\label{sec: Fan-Spine Topology near Sunspot and Magnetic-Reconnection behind splitting}


  NFFF extrapolations revealed that a fan-spine magnetic topology exists in the north-west direction near the sunspot, as displayed in \autoref{fig:magextrapolations}. Panel (B) of \autoref{fig:rcAIA171} shows a transient brightening observed in AIA 171 images between 19:50:23 UT and 19:56:35 UT. This brightening is also observed in hotter AIA channels, as depicted in Panel(A) of \autoref{fig:rcAIA171}, implying the existence of multi-thermal plasma in that location. A composite image of the brightening seen in AIA 171 at 19:55:11 UT with the NFFF-extrapolated field lines, as shown in panel (C) of \autoref{fig:magextrapolations}, indicates that the location of the brightening is near the footpoint of the fan-spine structure. Following this, multiple brightenings occur in this region, leading to a microflare producing a small-scale eruption around 20:58 UT (refer to the animation associated with Fig. \ref{fig:rcAIA171}). Considering this observational evidence, we conjecture that, due to a reconnection between closed field lines of fan-spine topology and nearby ambient open field lines, gradual evolution is occurring in the coronal loops associated with the downflow region. Note that a similar type of reconnection related to Type III coronal rain has been described in \citet{2024_Qiao_3Types_Coronal_Rain} albeit for an on-disk scenario. However, it is important to note that our NFFF extrapolations do not reveal any field lines that connect the region of the downflows directly to the nearby fan-spine structure (see \autoref{fig:magextrapolations}). This absence may indicate that the two regions are not directly magnetically connected, or it may also be attributed to low-resolution HMI data. Therefore, the causal link between the downflows and the microflare region is not evident at this moment.

\section{Discussions and conclusions} 
\label{sec:Discussions}
 In this article, we primarily study the spatio-temporal dynamics of a long-lived TR downflow above a sunspot (AR 12135) by utilising nearly 3-hour-long repeated-raster mode IRIS observations. After thoroughly examining the changes in spectral shape over time, instead of the traditional double Gaussian fitting, we performed triple Gaussian fitting with the same set of constraints on the TR spectral lines (Si IV 1394 and 1402, O IV 1399 and 1401) at each pixel to extract spectral parameters to study the spatial morphology and temporal evolution of this event. High-speed localised downflows are initially observed in both the umbral and penumbral regions. Over time, the penumbral downflow fades, while the umbral downflow strengthens and appears as a single structure that eventually splits into two spatial segments. At the end of the observation, only one segment remains strong enough to be detected in different spectral lines. 

We have performed NFFF extrapolation using a vector magnetogram from SDO/HMI at 19:36:00 UT to explore the underlying magnetic connectivity in that region. The results revealed a fan-spine magnetic topology near the sunspot. Additionally, sudden brightening is observed in AIA 171 images from 19:50:23 UT to 19:56:35 UT. Notably, these brightening events are recurring, and also visible in the hotter AIA channels (indicating the presence of multi-thermal plasma), eventually producing a microflare in that region. Based on this observational evidence, we hypothesise that magnetic reconnection between open magnetic field lines and closed fan-spine topology \citep{2024_Qiao_3Types_Coronal_Rain} is dictating the overall evolution in the loops.

The calculated average properties, including downflow speed, electron density, and mass flux for this event, are consistent with established studies (see \autoref{ssds_properties_table}). In particular, the downflow speed appears to gradually increase over time followed by a marginal decrease indicating an acceleration and deceleration in the downflow. While the accelerating behavior was previously reported \citep{Chitta2016, 2020_Nelson_SSD, 2020_Nelson_raster_downflows}, the decelerating phase is captured here for the first time. DEM analysis using AIA data by employing the sparse inversion method indicates that the TR downflow speed and emission from cool coronal plasma ($\log_{10}[T/\mathrm{K}] \in [5.5, 5.6]$) are showing similar evolution. At this stage, it remains unclear what drives the acceleration and deceleration of these TR downflows, as currently there are a limited number of studies available.

By combining IRIS spectroscopic observations with imaging data from different AIA filters, our analysis revealed that changes in the downflow's morphology are closely linked to alterations in the coronal loops' structure, consistent with the conclusions of previous research \citep{Chitta2016, 2020_Nelson_SSD, 2020_Nelson_raster_downflows}. In this respect, the possibility of siphon flows as a source cannot be completely disregarded, but the dynamic evolution of the coronal loops and the corresponding changes in the downflow morphology in our observed event indicate towards a coronal origin, at least in the present case. Additional analysis with more information from the coronal region is required to pinpoint the exact source for this event. Furthermore, in order to assess how common the observed downflow's behavior is, and identify their relation to the evolution of overlying coronal loops, a statistical study with a larger sample size from IRIS, alongside a future mission like Multi-slit Solar Explorer (MUSE) \citep{2022_MUSE_I, 2022_MUSE_II}, which can provide the corresponding spectral evolution in the corona is required. Such a study may also provide a deeper insight into the origin of these sunspot-specific flows.


\begin{acknowledgments}
We thank the anonymous referee for useful comments. We would like to thank Dr. Tanmoy Samanta for the fruitful discussion that greatly enriched our work. IRIS is a NASA Small Explorer mission developed and operated by LMSAL with mission operations executed at NASA Ames Research Center and major contributions to downlink communications funded by ESA and the Norwegian Space Center. AIA and HMI are instruments onboard the Solar Dynamics Observatory, a mission for NASA’s Living With a Star program. CHIANTI is a collaborative project involving George Mason University, the University of Michigan (USA) and the University of Cambridge (UK).  Financial support for BM is generously provided by the Department of Science and Technology (DST), Government of India, through the Aryabhatta Research Institute of Observational Sciences (ARIES). BM also acknowledges support from the AAS/SPD Thomas Metcalf Travel Award to attend Coronal Loop Workshop XII, where valuable feedback from attending experts significantly improved the final draft of the manuscript. Additionally, BM extends heartfelt thanks to Srinjana Routh for all the insightful discussions and valuable suggestions throughout this work. SKP is grateful to ISRO for a RESPOND grant (No. ISRO/RES/2/453/25-26) and to the Royal Society for the International Exchanges grant (No. IES/R3/243262). 
\end{acknowledgments}





\bibliography{ssds_citation}{}
\bibliographystyle{aasjournalv7}

\end{document}